\documentclass[smallextended]{svjour3}

\smartqed 

\usepackage{graphicx,subfigure,enumerate,pstricks,latexsym}
\usepackage[breaklinks=true,colorlinks=true,linkcolor=blue,urlcolor=blue,citecolor=blue]{hyperref}
\usepackage[english]{babel}

\def\beq{\begin{equation}} \def\eeq{\end{equation}}
\def\bea{\begin{eqnarray}} \def\eea{\end{eqnarray}}

\def\nn{\nonumber}

\newcommand{\nnd}{\discretionary{--}{--}{--}} 
\newcommand{\Schw}{Schwarz\-schild} \newcommand{\dS}{de~Sitter}
\def\linka{ --- }

\def\SS{\Sigma}

  \def\p{P}
\def\GG{G} \def\RS{R}  \def\der{|}

    \def\mir{\mathrm{r}} \def\mit{\mathrm{\theta}}

\def\tom{\tilde{\omega}}

\def\picA{10cm} \def\picB{5cm}

\begin{document}

\title{Test of the string loop oscillation model using kHz quasiperiodic oscillations in a neutron star binary}
\titlerunning{String loop oscillation model in a neutron star source}

\author{ Z. Stuchl\'{\i}k \and M. Kolo\v{s} }

\institute{Z. Stuchl\'{\i}k \and M. Kolo\v{s} \at
          Institute of Physics, Faculty of Philosophy \& Science, 
					Silesian University in Opava, 
					Bezru\v{c}ovo n\'{a}m\v{e}st\'{i} 13, CZ-74601 Opava, Czech Republic\\
\email{zdenek.stuchlik@fpf.slu.cz}, \email{martin.kolos@fpf.slu.cz}
}

\date{Received: date / Accepted: date}

\journalname{General Relativity and Gravitation}

\maketitle

\begin{abstract}
The model of current-carrying string loop oscillations is tested to explain the special set of frequencies related to the high-frequency quasiperiodic oscillations (HF QPOs) observed recently in the low-mass X-ray binary XTE J1701-407 containing a neutron star. The external geometry of the neutron star is approximated by the Kerr geometry, introducing errors not exceeding $10~\%$ for slowly rotating massive neutron stars. The frequencies of the radial and vertical string loop oscillations are then governed by the mass $M$ and dimensionless spin $a$ of the neutron star, and by the dimensionless parameter $\omega$ describing combined effects of the string loop tension and its angular momentum. It is explicitly demonstrated that the string-loop oscillation model can explain the observed kHz frequencies for the neutron star parameters restricted to the intervals ${0.2<a<0.4}$ and ${2.1<M/{\rm M}_{\odot}<2.5}$. However, the stringy parameter $\omega$ cannot be the same for all the three HF QPO observations in the XTE J1701-407 source; the limits on the acceptable values of $\omega$ are given in dependence on the spacetime parameters $M$ and $a$. 
\end{abstract}

\section{Introduction}\label{intro}

The axisymmetric current-carrying string loops are governed by their tension and angular momentum. Tension of the string loops prevents their expansion beyond some radius, while their worldsheet current introduces an angular momentum preventing them from collapse. Originally, the cosmic strings were introduced as remnants of the phase transitions in the very early universe \cite{Vil-She:1994:CSTD:}, or strings represented as superconducting vortices were considered in \cite{Wit:1985:NuclPhysB:}. However, the current-carrying string loops could represent also plasma exhibiting a string-like behaviour due to dynamics of the magnetic field lines \cite{Sem-Dya-Pun:2004:Sci:,Chri-Hin:1999:PhRvD:}, or due to the thin flux tubes of magnetized plasma simply described as 1D strings \cite{Sem-Ber:1990:ASS:,Cre-Stu:2013:PHYSRE:,Cre-Stu-Tes:2013:PlasmaPhys:,Kov:2013:EPJP:}.

In recent studies it has been demonstrated that the current-carrying string loops moving axisymmetrically along the symmetry axis of the Kerr or \Schw\nnd\dS{} black holes can have interesting direct astrophysical applications \cite{Jac-Sot:2009:PHYSR4:,Kol-Stu:2010:PHYSR4:,Stu-Kol:2012:PHYSR4:,Kol-Stu:2013:PHYSR4:}. For example, the transmutation effect related to the current-carrying string loops, i.e., transmission of their oscillatory internal energy into energy of the translational motion, causes an outward-directed acceleration of the string loops in the gravitational field of stars or compact objects, as neutron stars, black holes, or naked singularities \cite{Jac-Sot:2009:PHYSR4:,Stu-Kol:2012:PHYSR4:,Stu-Kol:2012:JCAP:,Kol-Stu:2013:PHYSR4:}. Such an effect can be important also for the electrically charged string loops moving in combined external gravitational and electromagnetic fields \cite{Tur-etal:2013:PHYSR4:,Tur-etal:2014:PHYSR4:}. 

Since acceleration of the string loops in the deep gravitational field of black holes or naked singularities can be extremely efficient, leading to ultra-relativistic escaping velocities of the string loops \cite{Stu-Kol:2012:JCAP:,Kol-Stu:2013:PHYSR4:}, the string loop transmutation can serve as a model of the formation of the ultra-relativistic jets observed in active galactic nuclei or Galactic microquasars. The string loop transmutation model of collimated relativistic jets can thus be considered as an alternative to the standard model based on the Blandford-Znajek process \cite{Bla-Zna:1977:MNRAS:}, or recently proposed model of geodesic collimation of ultra-relativistic particles \cite{Gar-etal:2010:ASTRA:,Gar-Mar-San:2013:ApJ:}. For acceleration of jets escaping from the active galactic nuclei the cosmic repulsion can be also relevant behind the so called static radius \cite{Stu-Kol:2012:PHYSR4:,Stu:1983:BULAI:,Stu-Hle:1999:PHYSR4:}. 

Quite recently, it has been demonstrated that the string loops can be astrophysically relevant in situation inverse to the transmutation effect. Small oscillations of the string loops around stable equilibrium positions in the equatorial plane of the Kerr geometry can be considered in the lowest approximation as two uncoupled linear harmonic oscillators governing the radial and vertical oscillations of the string loop -- the higher-order terms determine non-linear phenomena and subsequent transition to the quasi-periodic and chaotic oscillatory motion \cite{Kol-Stu:2013:PHYSR4:}. The frequencies of the radial and vertical harmonic oscillations of the string loops are relevant also in the quasi-periodic stages of the motion and their radial profiles were given and discussed in \cite{Stu-Kol:2014:PHYSR4:}. It has been shown that frequencies of the string loop harmonic or quasi-harmonic oscillations can fit twin HF QPOs observed with frequency ratio $3:2$ in three Galactic microquasars GRS~1915+105, XTE~1550-564, GRO~1655-40, i.e, low-mass X-ray binary (LMXB) systems containing a black hole \cite{Stu-Kol:2014:PHYSR4:}. 

The radial profiles of the string loop oscillations qualitatively differ from those related to the radial and vertical oscillations of the geodesic, test particle motion in the Kerr geometry, especially there is a crossing point of the radial and vertical frequencies in the Kerr black hole spacetimes for the string loop oscillations, while for the test particle oscillations such a crossing is possible only in the Kerr naked singularity spacetimes \cite{Tor-Stu:2005:ASTRA:,Stu-Sche:2012:CLAQG:}. This property of the radial profiles of the radial and vertical frequencies of the string loop oscillations is applied here to test the string loop oscillation model \cite{Stu-Kol:2014:PHYSR4:} for fitting the special frequency set of HF QPOs observed recently in the source XTE J1701-407 that is a LMXB system containing a neutron star \cite{Paw-etal:2013:MONRAS:}. For simplicity, we assume that the external geometry of the neutron star can be appropriately approximated by the Kerr geometry. 

\section{HF QPOs observed in the XTE~J1701-407 source}

The LMXBs can contain a black hole or a neutron star. The neutron star LMXBs are classified as {\it Z} sources if they demonstrate high luminosities ($0.5-1.0$ of Eddington luminosity $L_{\rm Edd}$), and {\it atoll} sources if they demonstrate low luminosities ($0.01-0.5~L_{\rm Edd}$) \cite{Has-vdK:1989:ASTRA:,vdK:2006:CompStell:}. 

The source XTE~J1701-407 has been discovered by the Rossi X-ray Timing Explorer (RXTE) at June 2008 \cite{Mar-etal:2008:AstrTel:}. Since thermonuclear X-ray bursts were observed at the XTE~J1701-407 source \cite{Chen-etal:2010:AstrTel:}, we can conclude that it contains an accreting neutron star. The discovery of HF QPOs at the XTE J1701-407 source was first reported in \cite{Str-etal:2008:AstrTel:} where it was mentioned that it is one of the least luminous atoll sources, having the X-ray luminosity $L_{\rm X} \sim 0.01 L_{\rm Edd}$. Later, it has been discovered that the source is a peculiar neutron star LMXB displaying all various types of the neutron star LMXB subclasses, i.e., Z-class, bright atoll and atoll \cite{Hom-etal:2010:ApJ:}

Recently, a detailed analysis of the HF QPOs in the source XTE~J1701-407 has been reported in \cite{Paw-etal:2013:MONRAS:}, being based on the whole ansamble of 58 RXTE observations. The results are rather unexpected and very interesting, since a very special set of frequencies has been discovered at three of the whole ansamble of the RXTE observations. In one of the three observational events a single HF QPO has been detected at a characteristic frequency \cite{Paw-etal:2013:MONRAS:} 
\beq
 f_{\rm(A)L} = f_{\rm(A)U} = 1153\pm5 \,\,\mathrm{Hz}. \label{ff11}
\eeq
In the other two observations twin HF QPOs has been detected at characteristic frequencies \cite{Paw-etal:2013:MONRAS:}
\bea
 f_{\rm(B)L} = 740\pm5 \,\,\mathrm{Hz}, \quad f_{\rm(B)U} = 1112\pm17 \,\,\mathrm{Hz}, \label{ff32a}\\
 f_{\rm(C)L} = 740\pm11 \,\,\mathrm{Hz}, \quad f_{\rm(C)U} = 1098\pm5 \,\,\mathrm{Hz}, \label{ff32b}
\eea
where we use the index U for the upper and the index L for the lower of the twin frequencies observed simultaneously. The other details of the detected HF QPOs, namely the fractional rms amplitudes and the quality factors, are presented in \cite{Paw-etal:2013:MONRAS:}. We do not discuss them here, since we concentrate our attention on the observed frequencies only. 

During the observations when the one-peak HF QPO has been detected, a low-frequency QPO has been also detected at $f_{low} = 30.4\pm0.3$~Hz \cite{Paw-etal:2013:MONRAS:} that is not considered in the following as we restrict our attention to the HF QPOs. However, we plan to study the possibility that the low-frequency QPO could be related to "rocking" of a string loop around the symmetry axis. 

Note that the HF (kHz) QPOs do not always appear as twin peaks, but only one peak is sometimes observed in the neutron star systems. If only one kHz peak is detected, it could mean that the other peak is not detected for physical reasons, or because of limitations of detectors, such as low signal-to-noise. The physical reasons could block production of the second peak, or, as we assume here, frequencies of the two oscillatory modes coincide. 

The frequency set corresponding to the HF QPOs detected in the XTE~J1701-407 source is really of a special character among the HF QPOs observed in the atoll sources, since a single HF QPO is combined with two twin HF QPOs having the frequency ratio $3:2$ typical for microquasars. We shall use the special character of the radial profiles of the frequencies of the harmonic radial and vertical oscillations of axially symmetric string loops, namely the coincidence of these frequencies at a special radius \cite{Stu-Kol:2014:PHYSR4:}, in order to fit quite naturally the single HF QPO observed in the source XTE~J1701-407. 

\section{Neutron stars and Kerr geometry}

In our test of the string loop model for the case of the neutron star system XTE~1701-407 we apply the Kerr geometry to describe the external geometry of the neutron star, as the radial profiles of the radial and vertical string loop oscillation frequencies are given by analytic formulae in this geometry and enable relatively simple and elegant fitting procedure. However, we have to clear up the conditions enabling using of such a strong simplification for the external geometry of the neutron star. 

\subsection{Hartle-Thorne theory of rotating neutron stars}

The rotating neutron stars can be conveniently described by the Hartle-Thorne geometry that is governed in its exterior part by three parameters: mass $M$, dimensionless spin $a$, and dimensionless quadrupole moment $q$. The Hartle-Thorne theory can be applied for rotating neutron stars with rotation frequency smaller than the mass-shedding frequency ($f_{\rm mass~shedding}\sim1100~\mathrm{Hz}$) and dimensionless spin $a < 0.4$ \cite{Har:1967:ApJ:,Har-Tho:1968:ApJ:,Cha-Mil:1974:MONNR:,Mil:1977:MONNR:,Urb-Mil-Stu:2013:MONRAS:}. For $a<0.4$ and $q/a^2 = 1$, the external Hartle-Thorne geometry \cite{Har-Tho:1968:ApJ:} coincides with the Kerr geometry with precision on the level of $\sim~1~\%$. For $1 < q/a^2 < 3$ these two geometries have very similar radial profiles of the metric coefficients, giving thus for astrophysical phenomena occurring in vicinity of such rotating neutron stars predictions with errors on the level lower than $\sim~5~\%$ \cite{Tor-Bak-Stu-Cech:2008:AcA:,Tor-etal:2010:ApJ:,Tor-etal:2012:ApJ:,Bej-Zdu-Hae:2010:ASTRA:,Bin-etal:2013:CLAQG:,Tor-etal:2014:ASTRA:}. 

Recently a very important result has been obtained in \cite{Urb-Mil-Stu:2013:MONRAS:} where it has been demonstrated that for the whole variety of considered realistic equations of state enabling existence of massive neutron stars with $M>2{\rm M}_{\odot}$, the Hartle-Thorne theory predicts $q/a^2<2$, if the mass of the neutron star with a given rotation frequency is close to the maximum allowed for a given equation of state. We can see in Figure 3 of \cite{Urb-Mil-Stu:2013:MONRAS:} that near the mass-maximum given by equations of state allowing for $M_{\rm max}>2{\rm M}_{\odot}$, the value of the characteristic parameter can be as low as $q/a^2\sim 1.5$. Therefore, such near-maximum-mass neutron stars can be properly described by the Kerr black hole geometry. The precision of the agreement of the external Hartle--Thorne and Kerr geometries is sufficiently high (with relative difference lower than $3~\%$), if the conditions $a~<~0.4$ and $q/a^2~\leq~2$ are satisfied \cite{Tor-etal:2010:ApJ:,Urb-Mil-Stu:2013:MONRAS:}. 

For masses substantially lower than the maximum allowed by the realistic equations of state, the parameter $q/a^2$ can be relatively high -- for $M \sim 0.7~M_{\rm max}$ there is $q/a^2 \to (5-10)$ in dependence on a concrete equation of state \cite{Urb-Mil-Stu:2013:MONRAS:}. We can see in Figure 3 of \cite{Urb-Mil-Stu:2013:MONRAS:} that for the so called "canonical" neutron star mass $M=1.4{\rm M}_{\odot}$, there is $q/a^2 \to (3-7)$ for all of the considered realistic equations of state. In such cases, dependence of astrophysical phenomena on the neutron star parameters $M,a,q$ can be very complex as demonstrated in \cite{Tor-etal:2014:ASTRA:} for the case of the position of the innermost stable circular geodesic ($r_{\rm ISCO}$) in the Hartle-Thorne spacetime. The difference of the $r_{\rm ISCO}$ in the Hartle-Thorne and Kerr spacetimes having identical parameters $M$ and $a$ reaches tens of percent and the dependence of $r_{\rm ISCO}$ on the spacetime parameters differs even qualitatively for any value of spin $a<0.4$ and for large values of the parameter $q$ giving $q/a^2 > 3$. On the other hand, the relative differences are less than $10 \%$ for spin $a<0.4$ and $q/a^2 < 3$ -- see Figure 1 in \cite{Tor-etal:2014:ASTRA:}. 

We can summarize that for the near-maximum mass neutron stars the situation is common and relatively simple for all equations of state if the spin parameter $a<0.4$, since the quadrupole moment parameter satisfies the condition $q/a^2 < 2$ implying similarity of the Hartle-Thorne and Kerr geometries and similar character of astrophysical phenomena in these spacetimes, with relative differences smaller than $5~\%$. The Hartle-Thorne metric is quadratic in the dimensionless spin $a$, and for $a \sim 0.4$ it introduces truncation errors of the order $0(a^3) \sim 6~\%$ implying total errors introduced by assumption of the Kerr background geometry about $10~\%$. Therefore, more detailed and accurate analysis of the oscillatory motion, based on the Hartle-Thorne model or numerical codes of fast-rotating neutron stars (RNS \cite{RNS,Kom-Eri-Hach:1989:MONNRS:} or LORENE \cite{LORENE,Bon-Gou-Mar:1998:PHYSR4:} codes) will be necessary in future studies. For masses significantly smaller than the maximal mass related to a given equation of state, the Hartle-Thorne geometry depends strongly on the equation of state having the parameter $q/a^2>3$; in such a case it differs significantly (qualitatively) from the Kerr geometry. Approximation of a neutron star exterior by the Kerr geometry is irrelevant in such situations. 

\subsection{Theoretical restrictions on the mass and spin of neutron stars}

Predictions of the string-loop oscillation model for the XTE~J1701-407 source containing a neutron star have to be confronted with the theoretical constraints on mass and spin of neutron stars. The upper theoretical constraint on the~neutron star mass based on realistic equations of state of the subnuclear matter reads
\beq
    M < M_{\rm maxNS} = 2.5\,\mathrm{M}_{\odot} \label{limit-na-M-z-rotace}
\eeq
\cite{Akm-Pan:1997:,Akm-Pan:1998:,Cham-Hae-Zdu-Fam:2013:IJMPE:}. Note that the limit governed by the causality condition gives $M_{\rm maxNS}\sim 3{\rm M}_{\odot}$ \cite{Cham-Hae-Zdu-Fam:2013:IJMPE:}, and the equation of state based on the relativistic mean-field theory \cite{Bodmer:1971:PHYSR4:} allows for $M_{\rm maxNS} \sim 2.8 {\rm M}_{\odot}$. 

The upper theoretical limit on the neutron star spin, based on the fully general relativistic set of numerical codes for numerical relativity, LORENE \cite{LORENE}, reads $a_{\rm maxNS} \sim 0.7$, as demonstrated in \cite{Lo-Lin:2011:ApJ:}, and corresponds to the maximal rotational frequency of the neutron stars, $f_{\rm rot}=f_{\rm mass~shedding}$. For $f_{\rm rot} \sim 0.6 f_{\rm mass~shedding}$ corresponding to the observed neutron stars with fastest rotation, the LORENE codes predict the neutron star spin $a \leq 0.4$ -- see Figure 3 in \cite{Lo-Lin:2011:ApJ:}. Surprisingly, the numerically obtained fully general relativistic, fast rotating, "LORENE" external geometry, and the Hartle-Thorne external geometry obtained in the "slow rotation" approximation of the neutron star models, differ for the equations of state considered in \cite{Urb-Mil-Stu:2013:MONRAS:} maximally on the level of $5\%$, if the near-maximal mass neutron stars with the dimensionless spin $a<0.4$ are considered. Their difference is substantially higher, $>10\%$, for the spin $a > 0.5$. Therefore, we put the theoretical limit on applicability of the Hartle-Thorne model of neutron stars to be 
\beq
     a < 0.4\,. \label{limit-na-spin-z-rotace}
\eeq

For quark (strange) stars, the~maximal mass constraint is smaller because of softer equations of state assumed in modelling the~strange stars, but masses around $M_{\rm maxQ} \sim 2\,\mathrm{M}_{\odot}$ are allowed \cite{Gle:2000:CompactStars:,Lo-Lin:2011:ApJ:}; in some special conditions the maximum mass can be substantially higher, namely $M_{\rm maxQ} \sim 2.7 {\rm M}_{\odot}$ \cite{Cham-Hae-Zdu-Fam:2013:IJMPE:}. On the other hand, slightly superspinning states of strange quark stars with $a_{\rm maxQ} \geq 1$, exceeding the~black hole limit $a=1$, are allowed as reported in \cite{Lo-Lin:2011:ApJ:}. Such a difference between the~maximal spin of neutron stars and strange stars can be explained by the~strong nuclear force acting in binding the~strange stars. Here we consider only the neutron star models. 

We use the theoretical restrictions (\ref{limit-na-M-z-rotace}) and (\ref{limit-na-spin-z-rotace}) as limits on the mass and spin of the XTE~J1701-407 neutron star, given by the string loop oscillation model applied to fit the frequencies of the HF QPOs in the three observational events reported in \cite{Paw-etal:2013:MONRAS:}. 

\section{Oscillations of current-carrying string loops}

An axisymmetric current-carrying string loop in a given axially symmetric and stationary spacetime with metric $g_{\alpha\beta}$ has been discussed in detail in \cite{Jac-Sot:2009:PHYSR4:,Kol-Stu:2013:PHYSR4:}. Oscillations of such string loops have been studied in \cite{Stu-Kol:2014:PHYSR4:}. Here we give a short overview. 

\subsection{Dynamics of axisymmetric string loops}

The worldsheet of a string loop is given by the spacetime coordinates $X^{\alpha}(\sigma^{a})$ with $\alpha = 0,1,2,3$ that are functions of two worldsheet coordinates $\sigma^{a}$ with $a = 0,1$. The induced metric on the worldsheet reads $h_{ab}= g_{\alpha\beta}X^\alpha_{\der a}X^\beta_{\der b}$ where $\Box_{\der a} = \partial \Box /\partial a$. The string loop current is governed by a scalar field $\varphi({\sigma^a})$. Dynamics of the string loop is determined by an action with Lagrangian $\mathcal{L} = -(\mu + h^{ab} \varphi_{\der a}\varphi_{\der b})\sqrt{-h}$ where $ \varphi_{,a} = j_a $ determines the current of the string and $\mu > 0$ reflects the string tension \cite{Vil-She:1994:CSTD:,Wit:1985:NuclPhysB:,Jac-Sot:2009:PHYSR4:,Kol-Stu:2010:PHYSR4:}. Axisymmetry of the string loop in accord with axisymmetry of the spacetime means that the scalar field $\varphi = j_{\sigma}\sigma + j_{\tau}\tau$, where $j_{\sigma}$ and $j_{\tau}$ are constant components of the current and represent the angular momentum of the string loop. We define the angular momentum parameters by the relations $J^2 = j_{\sigma}^2 + j_{\tau}^2$ and $\omega = - j_{\sigma}/j_{\tau}$ \cite{Kol-Stu:2013:PHYSR4:}. 

Dynamics of the axisymmetric string loops is then given by the worldsheet stress-energy tensor expressed through the tension $\mu > 0$ and the parameters $J>0$ and $\omega\in\langle-1,1\rangle$ that are related to the current on the string loop and determine its angular momentum. For a given string loop, the tension and both the angular momentum parameters are constants of the motion.\footnote{For an alternate approach to the modelling of strings see the vorton model of \cite{Car-Pet-Gan:1997:PHYSR4:,Car-Pet:1999:PRB:}, or strings with time-varying tension \cite{Wang-Cheng:2012:PhysLetB:}. A detailed information and review can be found in \cite{Vil-She:1994:CSTD:}.} The worldsheet stress-energy tensor of the string loop $\SS^{ab}$ reads (see \cite{Stu-Kol:2014:PHYSR4:} for details) 
\beq
 \SS^{\tau\tau} = \frac{J^2}{g_{\phi\phi}} + \mu , \quad \SS^{\sigma\sigma} = \frac{J^2}{g_{\phi\phi}} - \mu , \quad \SS^{\sigma\tau}  = \frac{2 \omega J^2}{g_{\phi\phi} (1+\omega^2)}. \label{sigma1}
\eeq

The string loop momentum components are given in the standard way as $P_\mu \equiv \frac{\partial \mathcal{L}}{\partial \dot{X}^\mu}$ where dot means derivative with respect to the proper time $\tau$. Dynamics of the string loops is then governed by the Hamilton equations \cite{Lar:1993:CLAQG:,Lar:1994:CLAQG:,Fro-Lar:1999:CLAQG:} with the Hamiltonian given in the form \cite{Lar:1993:CLAQG:,Kol-Stu:2013:PHYSR4:}
\beq
 H = \frac{1}{2} g^{\alpha\beta} \p_\alpha \p_\beta + \frac{1}{2} g_{\phi\phi} \left[(\SS^{\tau\tau})^2 - (\SS^{\tau\sigma})^2 \right] \label{AllHam} .
\eeq

The equations of evolution of axisymmetric string loops are similar to the equations governing evolution of toroidal magnetic fields and plasmas in a given background \cite{Sem-Dya-Pun:2004:Sci:,Chri-Hin:1999:PhRvD:,Jac-Sot:2009:PHYSR4:,Kol-Stu:2010:PHYSR4:}. Such composites of toroidal magnetic fields nad plasmas can be generated by the kinetic dynamo effect \cite{Cre-Stu:2013:PHYSRE:}. The oscillating string loops can represent a simplification of oscillating toroidal structures of magnetized plasmas, reflecting by their tension the effect of trapped magnetic toroidal field, while their angular momentum reflects the role of the matter content of such toroidal configurations. The oscillating string loop model can be thus considered as complementary to models of oscillating perfect-fluid tori reflecting the role of pressure gradients \cite{Rez-Yos-Zan:2003:MONNRS:,Str-Sra:2009:CLAQG:} where the effect of tension is not taken into account. We expect that physical parameters of a thin flux tube of magnetized plasma can be approximated by the string loop parameters $J,\omega$. 

\subsection{Energy boundary function}

The stationarity and the axial symmetry assumed for the spacetime imply existence of two conserved quantities - the string-loop energy $E$ and axial angular momentum $L$ that are given by 
\beq
 - E = P_t, \quad  L = P_\phi = g_{\phi\phi} \SS^{\sigma\tau} = -2j_{\tau}j_{\sigma}. \label{Cmotion}
\eeq
The axisymmetric current-carrying string loops are thus characterized by the total angular momentum $J$ and the axial angular momentum $L$. The case $\omega<0$ ($\omega>0$) represents string loops with negative (positive) axial angular momentum $L<0$ ($L>0$); $\omega=~0$ represents the string loops with $L=0$ \cite{Jac-Sot:2009:PHYSR4:,Kol-Stu:2013:PHYSR4:,Stu-Kol:2014:PHYSR4:}.

The Hamiltonian can be written as a sum of the dynamic and potential parts
\beq
 H = H_D + H_P = \frac{1}{2} g^{rr} \p_r^2 + \frac{1}{2} g^{\theta\theta} \p_\theta^2 + H_P(r,\theta).
\eeq 
Using the energy $E$ and the angular momentum parameters $J$ and $\omega$, the potential part of the Hamiltonian reads \cite{Stu-Kol:2014:PHYSR4:}
\beq
 H_{\rm P} =
 \frac{1}{2} g_{\phi\phi} \left( \frac{J^2}{g_{\phi\phi}}+1\right)^2
 + \frac{1}{2} \frac{g_{\phi\phi}}{\left( g_{tt} g_{\phi\phi}- g_{t\phi}^2\right)} \left( E +\frac{g_{t\phi}}{ g_{\phi\phi}} \frac{2 J^2 \omega }{\omega^2+1}\right)^2 .
	\label{HamHam}
\eeq
The boundary of the string loop motion is given by the condition $H_{\rm P}=0$ that implies the relation  \cite{Kol-Stu:2013:PHYSR4:,Stu-Kol:2014:PHYSR4:} 
\beq
 E = E_{\rm{b}}(r,\theta) = \sqrt{g_{t\phi}^2-g_{tt}g_{\phi\phi}} \, \SS^{\tau\tau} - g_{t\phi}\SS^{\sigma\tau} \label{EqEbRT} \label{StringEnergy}.
\eeq  
The energy boundary function $E_{\rm{b}}(r,\theta)$ governs the dynamics of the string loops, serving as an effective potential of their motion. 

We express the energy boundary function in the Kerr metric using the motion constants $J$ and $\omega$, making the rescaling $E_{\rm b} / \mu \rightarrow E_{\rm b}$ and $J / \sqrt{\mu} \rightarrow J$ -- such a choice of ``units'' is equivalent to setting the string tension $\mu=1$. The energy boundary function then, in the standard Boyer-Lindquist $r,\theta$ coordinates \cite{Car:1973:BlaHol:}, takes the form \cite{Stu-Kol:2014:PHYSR4:}  
\beq
 E_{\rm b} (r,\theta;a,J,\omega) = \frac{4 a \omega J^2 r }{\left(\omega^2+1\right) \GG }+\sqrt{\Delta}
   \left(\frac{J^2 R^2}{\GG \sin(\theta)}+\sin(\theta)\right), \label{EBrtKerr}
\eeq
with 
\beq    
  \GG (r,\theta;a) = \left(a^2+r^2\right) R^2 +2 a^2 r \sin^2(\theta ). \label{Gfce}
\eeq
The Kerr metric is governed by the characteristic functions 
\beq
\RS^2 = r^2 + a^2 \cos^2\theta, \quad \Delta = r^2 - 2Mr + a^2,
\eeq
where $a$ denotes spin and $M$ mass parameters of the Kerr spacetimes. In the following, we shall use for simplicity the dimensionless radial coordinate $r/M\rightarrow r$, dimensionless time coordinate $t/M\rightarrow t$ and dimensionless spin $a/M\rightarrow a$; this is equivalent to using of $M=1$ in the metric tensor. We will return to the dimensional quantities in the Section \ref{observations}.  

Detailed discussion of the properties of the energy boundary function $E_{b}(r,\theta)$ is presented in \cite{Kol-Stu:2013:PHYSR4:} for both the Kerr black hole and naked singularity spacetimes. Here we focus on its properties in the black hole spacetimes that can be relevant for rotating neutron stars as demonstrated in \cite{Urb-Mil-Stu:2013:MONRAS:,Tor-Bak-Stu-Cech:2008:AcA:} -- the local extrema of the energy boundary function can be then located in the equatorial plane only. 

The stationary points, i.e., the local extrema of the energy boundary function $E_{\rm b}(r;a,J,\omega)$, governing the equilibrium positions of the string loops in the equatorial plane ($\theta=\pi/2$), are determined by the function $J^2_{\rm E}(r;a,\omega)$ defined by the relation \cite{Kol-Stu:2013:PHYSR4:,Stu-Kol:2014:PHYSR4:}
\beq
J^2_{\rm E}(r;a,\omega) =  \frac{(r-1) \left(\omega^2+1\right) H^2}{4 a \omega \sqrt{\Delta} \left(a^2+3 r^2\right)+\left(\omega^2+1\right) F }, \label{eqrovina}
\eeq
where 
\beq
  H(r;a) = a^2 (r+2) +r^3, \quad F(r;a) = (r-3) r^4 -2 a^4+a^2 r \left(r^2-3 r+6\right). \label{Ffce}
\eeq

A detailed discussion of the properties of the energy boundary function $E_{\rm b}(r;a,J,\omega)$ and the string loop motion can be found in \cite{Kol-Stu:2013:PHYSR4:,Stu-Kol:2014:PHYSR4:}. We have to concentrate on the situation when for a string loop with fixed values of the angular momentum parameters $J$ and $\omega$ closed $E={\rm{}const}$ sections of the effective potential (energy boundary function) occur around a stable equilibrium position of the string loop given by the equation 
\bea
              J^2 = J^2_{\rm E}(r;a,\omega) .
\eea
The stable equilibrium position corresponds to the minimal energy related to the string loop with the angular momentum parameters $J,\omega$. Around such stable equilibrium positions, small oscillations of string loops occur, if their energy slightly exceeds the minimal value. The analysis of the oscillatory motion of string loops around their stable equilibrium positions, using the perturbative treatment of the Hamiltonian, can be found in \cite{Stu-Kol:2012:JCAP:,Kol-Stu:2013:PHYSR4:,Stu-Kol:2014:PHYSR4:}. 

\subsection{Frequency of the string-loop radial and vertical oscillatory modes}

We consider string loop harmonic oscillations around a stable equilibrium position with fixed coordinates $r_0$ and $\theta_0=\pi/2$ corresponding to the local minima of the energy boundary function. For the radial and latitudinal oscillatory string loop motion in the Kerr spacetimes, the angular frequencies related to distant observers are given by \cite{Stu-Kol:2014:PHYSR4:} 
\bea
 \Omega^2_{\mir}(r) &=& \frac{ J_{\rm E(ex)} \, \left[2 a \omega \sqrt{\Delta } \left(a^2+3 r^2\right)+ \tom \left(a^2 r^3 -a^2 \Delta  +r^5-2 r^4\right)\right]}{2 r H^2 \left[2a\omega  \left(a^2+3r^2\right) + \tom \sqrt{\Delta} \left(r^3-a^2\right)\right]^2} , \\
 \Omega^2_{\mit}(r) &=& 
\frac{a \left(2 \omega \sqrt{\Delta} - a \tom \right) \left[a^2 (2-3 r)-3 r^3\right] +r^2 \tom \left[a^2 (r-6)+r^3\right]}
{ r^2 H  \Delta^{-1/2} \left[2a\omega \left(a^2+3r^2\right) + \tom \sqrt{\Delta} \left(r^3-a^2\right)\right]}, 
\eea
where $\tom=\left(\omega ^2+1\right)$ and
\bea
 J_{\rm E(ex)} (r) &\equiv& H \tom (r-1) \left( 6 a^2 r -3 a^2 r^2 -6 a^2 -5 r^4+ 12 r^3 \right) \nn \\
 && + \tom \left[2 F (a^2+3r^2) (1-r) - F H \right] + 8 a \omega \sqrt{\Delta} (r-1) (a^2+3r^2)^2 \nn \\
 && + 4 a \omega \Delta^{-1/2} H \left[ (a^2+3r^2) \left(\Delta -(r-1)^2\right) -6 r \Delta (r-1) \right] . \label{JEex} 
\eea
The function $J_{\rm E(ex)}(r;a,\omega)$ governs the local extrema of the function $J_{\rm E}(r;a,\omega)$. Its zero points determine the marginally stable equilibrium positions of the string loops - at the zero points the frequency of the radial oscillatory modes of the string loops vanishes. \footnote{Exactly the same situation occurs in the case of the radial epicyclic motion of test particles \cite{Tor-Stu:2005:ASTRA:}.} The string-loop oscillations are possible only if the stable equilibrium positions of the string loops are allowed. The conditions 
\beq
  J_{\rm E(ex)}=0 \quad \textrm{and} \quad J_{\rm E}^2 \geq 0,
\eeq   
satisfied simultaneously, put the limit on validity of the formulae giving the angular frequencies of the radial and vertical oscillations - for details see \cite{Stu-Kol:2014:PHYSR4:}. 

\begin{figure}
\centering 
\includegraphics[height=\picA]{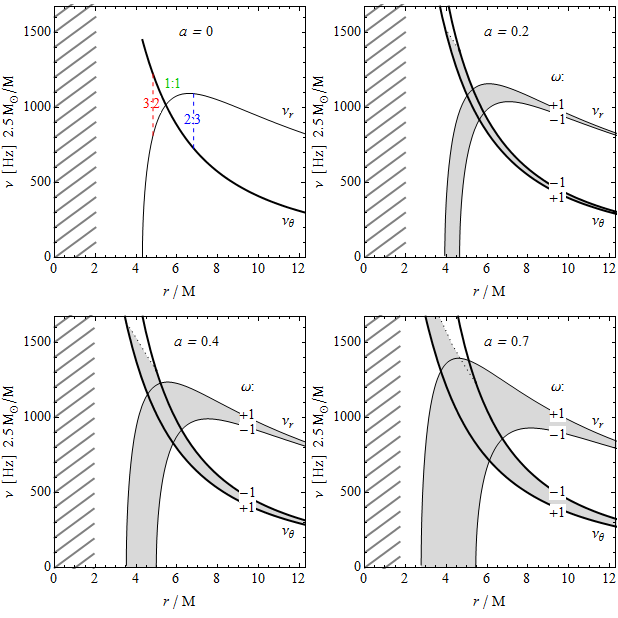}
\caption{\label{ffrange}
String-loop oscillatory frequencies $\nu_\mir$ (thin curves) and $\nu_\mit$ (thick curves), calculated for the Kerr metrics with $M = 2.5 {\rm M}_{\odot}$. Their radial profiles are illustrated for values of dimensionless spin $a=0,0.2,0.4,0.7$ that are characteristic of our study of neutron star system. We demonstrate extension of the frequency radial profiles for the complete range of the string loop parameter $\omega\in\langle-1,1\rangle$. The vertical frequency curves are restricted to the region of existence (zero point) of the corresponding radial frequency curves - the relevant region is greyed.
}
\end{figure}

\begin{figure}
\centering 
\includegraphics[height=\picB]{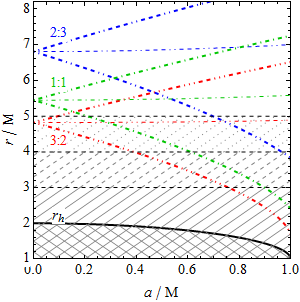}
\caption{ \label{RRfig}
The resonant radii where the radial and vertical frequencies of the string-loop oscillations have rational ratio $3:2$, $1:1$ and $2:3$, in dependence on the Kerr spin parameter $a$ and the complete range of the string-loop parameter $\omega$; the case of $\omega=0$ is given by thin lines, the upper thick lines are given for $\omega=-1$, the lower thick lines for $\omega=+1$. The loci of the outer horizon (thick) and radii $r=3,4,5$ (dashed) that represent possible radii of the neutron stars are also presented. Clearly, the string loop oscillations are allowed only above the surface of the neutron stars.}
\end{figure}

The radial profiles of the frequency of the radial and vertical modes of the string loop harmonic (or quasi-harmonic) oscillations has been discussed in \cite{Stu-Kol:2014:PHYSR4:}. A very complex situation occurs for the case of near-extreme Kerr black hole spacetimes with $a > 0.99$, similarly with the situation of the test particle and fluid motion related to the so called Aschenbach effect \cite{Asch:2004:ASTRA:,Stu-Sla-Tor-Abr:2005:PHYSR4:,Stu-Sla-Tor:2007a:ASTRA:}. In the Kerr spacetimes with dimensionless spin $a < 0.7$ which has been shown to be the upper limit on the spin of  neutron stars, independently of the equation of state \cite{Lo-Lin:2011:ApJ:}, the properties of the radial profiles of the string loop oscillations are much simpler, as demonstrated in Figure \ref{ffrange} for four characteristic values of the Kerr spin parameter $a = 0, 0.2, 0.4, 0.7$; in the \Schw{} spacetime ($a=0$), the situation is degenerated, since both the frequencies are independent of the parameter $\omega$. The range of the radii where the oscillating string loops can occur due to possible variance of the stringy parameter $\omega$ increases with increasing spacetime dimensionless spin parameter $a$, as illustrated in Figure \ref{RRfig}. 

\section{String-loop oscillations as a model of HF QPOs in the XTE J1701-407 source} 
\label{observations}

We realize fitting of the frequencies of the kHz QPOs observed in the neutron star source XTE~1701-407 using the analytical radial and vertical frequency relations of the string loop harmonic oscillations in the Kerr geometry. Our preliminary results \cite{Stu-Kol:2015:inpreparation:} demonstrate that the relative difference of the frequency relations in the Kerr geometry with $a<0.4$ and the Hartle-Thorne geometry with $a<0.4$ and $q/a^2<2$ do not exceed $5~\%$, giving thus sufficient accuracy to approve the application of the analytic Kerr formulae instead of complex, numerically treated Hartle-Thorne formulae. We also assume relevance of the resonant phenomena of the string loop oscillations at the radii where $3:2$ and $1:1$ frequency ratios occur. 

\begin{table}
\begin{center}
\begin{tabular}{c c c c c}
\hline
				& & $f_{\rm(A)}$ & $f_{\rm(B)}$ & $f_{\rm(C)}$ \\
\hline
case 1 & $\nu_\mir:\nu_\mit$ & 1:1 & 3:2 & 2:3 \\
case 2 & $\nu_\mir:\nu_\mit$ & 1:1 & 2:3 & 3:2 \\
case 3 & $\nu_\mir:\nu_\mit$ & 1:1 & 3:2 & 3:2 \\
case 4 & $\nu_\mir:\nu_\mit$ & 1:1 & 2:3 & 2:3 \\
\hline
\end{tabular}
\caption{ Four possible combinations of HF QPOs observed in the XTE J1701-407 source.} \label{tab0}
\end{center}
\end{table}

\subsection{Fitting observational frequencies by frequencies of the oscillatory modes of string loops}

\begin{figure}
\centering 
\includegraphics[width=\hsize]{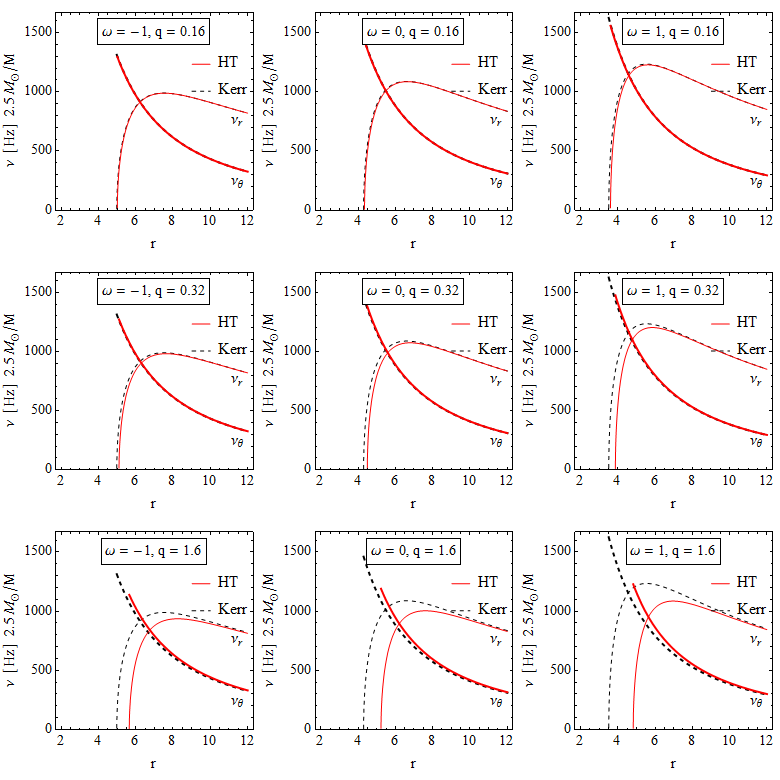}
\caption{ \label{HTfig}
Radial profiles of the string loop harmonic oscillation frequencies $\nu_{\mit}$ and $\nu_{\mir}$, related to the distant observers, are given for the Hartle-Thorne and Kerr spacetimes, and for the string loop parameter $\omega\in\{-1,0,1\}$. Both the Hartle-Thorne and Kerr spacetimes are rotating with the dimensionless spin $a=0.4$, while the quadrupole moment $q$ of the Hartle-Thorne spacetime takes the values $q\in\{0.16,0.32,1.6\}$.
For Hartle-Thorne spacetimes, the string loop oscillation frequencies are given only numerically.
In the case of $q=0.32$ ($q/a^2 = 2$), the relative difference $|\nu_{\rm Kerr} - \nu_{\rm HT}|/\nu_{\rm HT}$ of the Hartle-Thorne and the Kerr frequency radial profiles taken at the radius where the ratio $\nu_{\mit} : \nu_{\mir} = 1:1$, is not exceeding $5~\%$. 
}
\end{figure}

The frequencies of the string loop oscillations measured by the distant observers are in the dimensional form given by 
\beq
     \nu_{(\mir,\mit)} = \frac{1}{2\pi} \frac{c^3}{GM} \, \Omega_{(r,\theta)}.
\eeq
The frequencies of the harmonic string loop oscillations in the Kerr geometry have been compared to preliminary results of the analysis of the string loop oscillations in the Hartle-Thorne spacetimes \cite{Stu-Kol:2015:inpreparation:} in order to justify the application of the Kerr geometry. The comparison has been done for the cases where we expect small differences in the frequency radial profiles, namely for spin $a=0.4$ and the parameter $q/a^2=1$ and $q/a^2=2$. For comparison the case of $q/a^2=10$ is included. We can see in Figure \ref{HTfig} that the relative difference is really low for the cases $q/a^2=1$ and $q/a^2=2$, and high for $q/a^2=10$. For the neutron stars with near-maximal mass, dimensionless spin $a<0.4$, and low parameter $q/a^2 < 2$, the Kerr formulae for the harmonic frequencies $\nu_{\mit}$ and $\nu_{\mir}$ imply relative difference to the corresponding Hartle-Thorne formulae, $|\nu_{\rm Kerr} - \nu_{\rm HT}|/\nu_{\rm HT}$, lower than $5~\%$. 

In the LMXB systems containing a black hole or a neutron star, some of the~HF~QPOs come in pairs of the~upper and lower frequencies ($\nu_{\mathrm{U}}$, $\nu_{\mathrm{L}}$) of {\it twin peaks} in the~Fourier power spectra that are sometimes combined with individually detected HF QPOs. In the LMXB systems containing a black hole, so called microquasars, the twin HF QPOs occur at fixed frequencies that are usually at the exact $3:2$ ratio \cite{McC-etal:2011:CLAQG:}. On the other hand, in the LMXB systems containing a neutron star, large scatter of the twin HF QPOs is observed, and their frequency ratio varies across the frequency ratio $3:2$ down to $5:4$ ratio in the atoll sources \cite{Bar-Oli-Mil:2005:MONNRS:,Bel-Men-Hom:2007:MONNRS:,Tor-Bak-Stu-Cech:2008:AcA:,Wang-etal:2013:MONNRS:,Wang-etal:2014:AstrNach:}, while it varies in the range $4:1$ to $2:1$ in the Z-source Circinus~X1 \cite{Bout-etal:2006:ApJ:,Tor-etal:2010:ApJ:}. 

Usually, the Keplerian orbital and epicyclic (radial and latitudinal) frequencies of geodetical circular motion \cite{Ste-Vie:1999:PHYSRL:,Tor-Stu:2005:ASTRA:,Kot-Stu-Tor:2008:CLAQG:,Stu-Kot:2009:GenRelGrav:,Sche-Stu:2009:GenRelGrav:,Stu-Sche:2012:CLAQG:,Mot-etal:2014:MONRAS:,Bam-Mal-Tsu:2014:PHYSR4:} are assumed in models explaining the HF QPOs in both black hole and neutron star systems. Alternatively, the oscillations of tori \cite{Rez-Yos-Zan:2003:MONNRS:} or tilted oscillating discs \cite{Kato:2008:PASJ:} are considered as explanation of the HF QPOs. The frequencies of the oscillating tori and tilted discs are related to the orbital and epicyclic geodesic frequencies, if the oscillations are governed mainly by the gravity of the black hole or the neutron star \cite{Stu-Kot-Tor:2013:ASTRA:}. 

The assumption that resonances of oscillatory modes of Keplerian accretion discs have to be relevant seems to be confirmed due to the properties of the HF QPOs observed in the three Galactic microquasars GRS 1915+105, XTE 1550-564, GRO 1655-40, where the $3:2$ frequency ratio has been observed \cite{Klu-Abr:2000:ASTROPH:,Tor-etal:2005:ASTRA:}. However, the observed frequencies of the HF QPO in the three microquasars cannot be explained by a unique model based on the frequencies of the geodesic epicyclic motion \cite{Tor-etal:2011:ASTRA:,Bam:2012:JCAP:}. In the LMXBs containing neutron stars the fitting of the twin HF QPOs is realized by the relativistic precession model \cite{Ste-Vie:1998:ApJ:,Ste-Vie-Mor:1999:ApJ:} or its variations \cite{Tor-etal:2010:ApJ:}. However, the quality of the fits is not satisfactory for the models based on the geodesic orbital and epicyclic frequencies \cite{Tor-etal:2012:ApJ:}, although in some sources (4U 1636-53) the Resonant Switch model \cite{Stu-Kot-Tor:2012:AcA:,Stu-Kot-Tor:2013:ASTRA:} can improve the fits significantly without introducing a new free parameter in addition to the spacetime parameters \cite{Stu-etal:2014:ACTA:}.  

The quasi-periodic character of the oscillatory motion of current-carrying string loops trapped in a toroidal space around the equatorial plane of the central strong gravity object suggests application to the HF QPOs observed in the LMXB systems because of the same mass scaling of the frequencies of the radial and vertical oscillatory modes of string loops as those of the epicyclic geodesic frequencies, and comparable magnitudes of the frequencies of the string loop and epicyclic oscillations \cite{Stu-Kol:2014:PHYSR4:}. The string loop oscillation models has been successfully applied for the black hole systems in \cite{Stu-Kol:2014:PHYSR4:}, here we test the model for the special $kHz$ frequency set observed in the neutron star system XTE 1701-407. 

For a given twin HF QPOs observed in a given source, we have to consider fixed values of the string parameter $\omega$ and the spacetime parameters $M$ and $a$. If several twin HF QPOs are observed in the source, the spin and mass parameters have to be fixed, but the string loop parameter $\omega$ can be varied, as different twin frequency observations could be generated by different string loops that could be created and decayed successively with different parameter $\omega$ reflecting locally different conditions in the accretion discs of the source. Therefore, the string-loop oscillation model naturally introduces a possibility of significant scatter in distribution of frequencies of the twin HF QPOs. Such a scatter can occur because of shifting of the created string loops along the accretion disc, or due to varied conditions at a fixed radius of the disc. 
On the other hand, the resonance phenomena can be relevant for the behaviour of the oscillating string loops as indicated by the Kolmogorov-Arnold-Moser (KAM) theory \cite{Mos:1962:NAWG:,Stu-Kol:2014:PHYSR4:}, or some resonance phenomena could be relevant even for creation of the string loops, selecting thus some special radii related to the resonant phenomena.

\subsection{Fitting the kHz frequencies observed in XTE J1701-407}

The special set of the HF QPOs observed in the XTE~J1701-407 source indicates application of the string loop oscillation model assuming a resonance entering the play as in the microquasars. The presence of the single HF QPO observed in the XTE J1701-407 neutron star system can be well attributed to the fundamental property of the radial profiles of the string loop oscillatory radial and vertical modes, giving for each value of the stringy parameter $\omega$ a resonant position with frequency ratio $\nu_{\mir}:\nu_{\mit} = 1:1$. 

\begin{figure}
\centering 
\includegraphics[height=\picA]{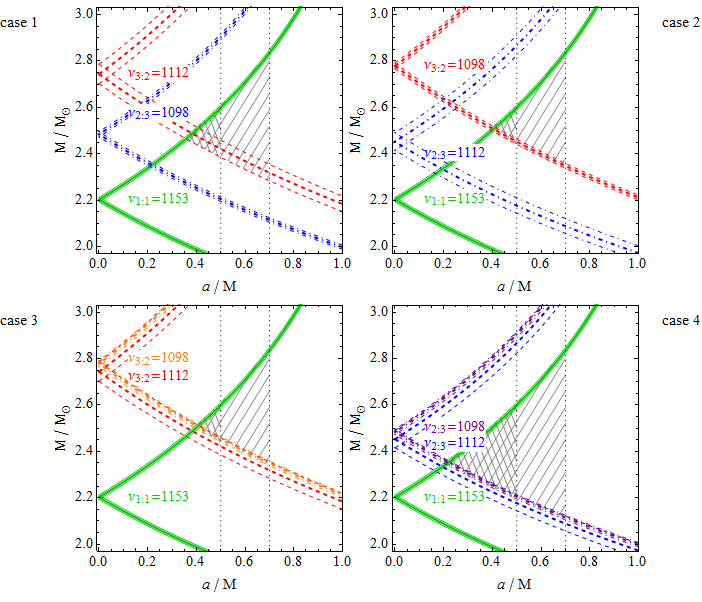}
\caption{\label{fits1}
Restrictions on the mass $M$ and spin $a$ parameters of the neutron star in the XTE~J1701-407 source implied by the string loop oscillation model applied to the three observational events of HF QPOs at the source. We assume that the three observational events occur at the resonant points of the radial and vertical string loop oscillations. Four different cases of the combinations of the resonant points related to the three observational events are possible -- see Table 1. In the fitting procedure related to the upper observed frequencies, we have included the mean values of the observed frequencies (thick) and both the lower and upper values given by measurement errors (thin), see (\ref{ff11}-\ref{ff32b}). The upper branches are for parameter $\omega=+1$ and the lower branches for parameter $\omega=-1$, allowed regions of the spacetime parameters $M,a$ are hatched. The most promising is the case 4, where we consider the $1:1$ resonance with frequency $\nu_{1:1}=1153$ Hz combined with two $2:3$ resonances with frequencies $\nu_{2:3} = 1088,1112$ Hz.
}
\end{figure}

\begin{table}[!h]
\begin{center}
\begin{tabular}{l c c c}
case 1: & & & \\
\hline
 & $a=0.350$ & $a=0.5$ & $a=0.7$ \\
\hline \hline
$M/ \mathrm{M}_{\odot}$ & 2.47 & $2.38${\linka}$2.61$ & $2.28${\linka}$2.85$ \\
\hline
$\omega_{3:2}$ & -1    & -1{\linka}-0.27   & -1{\linka}0.03  \\
$\omega_{2:3}$ & -0.02 & -0.12{\linka}0.19 & -0.21{\linka}0.37 \\
$\omega_{1:1}$ & 1     & 0.30{\linka}1     &  0.11{\linka}1\\
\hline
\\
case 2: & & & \\
\hline
 & $a=0.389$ & $a=0.5$ & $a=0.7$ \\
\hline \hline
$M/ \mathrm{M}_{\odot}$ & 2.50 & $2.44${\linka}$2.61$ & $2.34${\linka}$2.85$ \\
\hline
$\omega_{3:2}$ & -1   & -1{\linka}-0.28  & -1{\linka}0.03  \\
$\omega_{2:3}$ & 0.11 & 0.06{\linka}0.19 & -0.06{\linka}0.38 \\
$\omega_{1:1}$ & 1    & 0.39{\linka}1    &  0.16{\linka}1\\
\hline
\\
case 3: & & & \\
\hline
 & $a=0.389$ & $a=0.5$ & $a=0.7$ \\
\hline \hline
$M/ \mathrm{M}_{\odot}$ & 2.50 & $2.44${\linka}$2.61$ & $2.34${\linka}$2.85$ \\
\hline
$\omega_{3:2}$ & -1    & -1{\linka}-0.28    & -1{\linka}0.03  \\
$\omega_{3:2}$ & -0.46 & -0.50{\linka}-0.27 & -0.55{\linka}0.03 \\
$\omega_{1:1}$ & 1     & 0.39{\linka}1      &  0.16{\linka}1\\
\hline
\\
case 4: & & & \\
\hline
 & $a=0.201$ & $a=0.5$ & $a=0.7$ \\
\hline \hline
$M/ \mathrm{M}_{\odot}$ & 2.35 & $2.19${\linka}$2.61$ & $2.11${\linka}$2.85$ \\
\hline
$\omega_{2:3}$ & -1    & -1{\linka}0.19    & -1{\linka}0.37  \\
$\omega_{2:3}$ & -0.54 & -0.47{\linka}0.19 & -0.52{\linka}0.38 \\
$\omega_{1:1}$ & 1     & 0.01{\linka}1     &  0.10{\linka}1\\
\hline
\end{tabular}
\caption{ 
Restriction on the parameters of the neutron star in the XTE~J1701-407 source implied by the string loop oscillation model of HF QPOs. Presented values correspond to the hatched regions from Fig. \ref{fits1}. 
} \label{tab1}
\end{center}
\end{table}

\begin{figure}
\centering 
\includegraphics[width=\hsize]{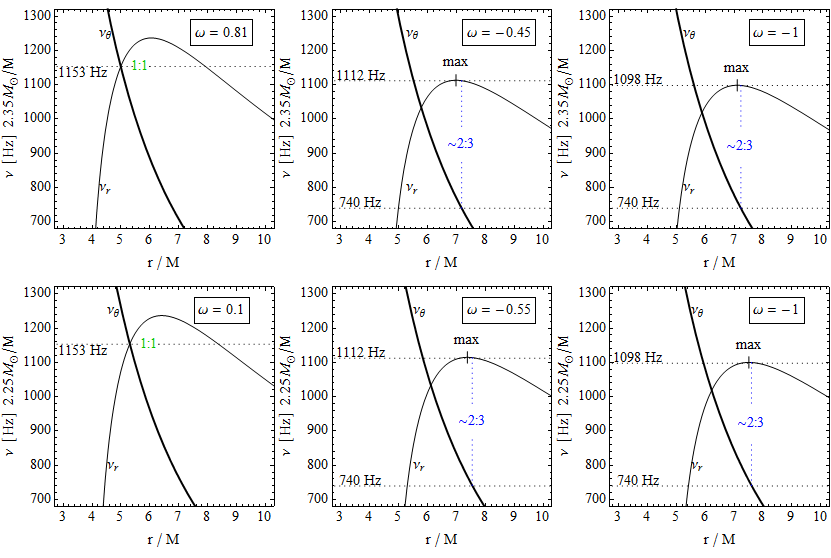}
\caption{ \label{rprofile}
Examples of the radial profiles of the string-loop oscillatory frequencies $\nu_\mir$ (thin curves) and $\nu_\mit$ (thick curves) as related to the three observational events are given for two representative situations allowed by the combination of the resonant points in the case 4. The parameters of the Kerr metric are mass $M = 2.35 {\rm M}_{\odot}$ and spin $a=0.22$ for the first row, $M = 2.25 {\rm M}_{\odot}$ and spin $a=0.4$ for the second row. The related values of the parameter $\omega$ are depicted in all the subfigures. Relevant resonant frequencies $\nu_{1:1} = 1153$~Hz and $\nu_{2:3} = (1110,740)$~Hz are given. The resonant radii $r_{2:3}$ are close to the maximum of the radial profile of the radial mode of the string loop oscillations.}
\end{figure}

We assume relevance of resonance phenomena, e.g., a parametric resonance \cite{Lan-Lif:1969:Mech:}, at all of the three HF QPO events observed in the XTE~J1701-407 source. We consider the rational frequency ratios $\nu_{\mit}:\nu_{\mir} = 3:2$ or $\nu_{\mit}:\nu_{\mir} = 2:3$, and $\nu_{\mit}:\nu_{\mir} = 1:1$, to be directly related to the observed values of the QPO frequencies in the XTE J1701-407 source. For all of the three observational events, we identify directly the frequencies $\nu_{\rm U}, \nu_{\rm L}$ with $\nu_{\mit}, \nu_{\mir}$ or $\nu_{\mir}, \nu_{\mit}$  frequencies. There are four possible combinations of this identification enabled by the properties of the string loop oscillation model and presented in Table 1. In the cases 1, 2, all three of the resonant radii of the string loop oscillations, $r_{3:2}$, $r_{1:1}$, $r_{2:3}$, are considered, while in the cases 3,4, the resonant radius $r_{1:1}$, and two resonant radii $r_{3:2}$ (case 3), or two resonant radii $r_{2:3}$ (case 4), are considered. 

In a given spacetime (fixed parameters $M,a$), the string loop oscillation model predicts for a string loop with a fixed parameter $\omega$ that the resonance radii and the radial (or vertical) frequencies related to the upper observed frequency have to satisfy the relations 
\beq
 r_{3:2} < r_{1:1} < r_{2:3}, \quad \nu_{3:2} > \nu_{2:3} > \nu_{1:1},
\eeq
where we denote as $\nu_{m:n}$ the upper frequency in the $m:n$ resonance point -- see Fig. \ref{ffrange}. However, in the XTE J1701-407 source, there is (eqs. (\ref{ff11}-\ref{ff32b})) 
\beq
 f_{1:1} > f_{3:2}, 
\eeq
therefore, it is impossible to explain all the observed resonances by assuming only one string loop with a fixed parameter $\omega$. We thus assume that at the resonant radii corresponding to the three observational events with HF QPOs, different string loops occur, being governed by different values of the parameter $\omega$. The string loop parameter $\omega$ is thus considered to be a free parameter in a given spacetime with fixed parameters $M,a$. 

The procedure of fitting the string loop oscillation frequencies to the observed frequencies in the three observational events is presented in Figure \ref{fits1} for all the four cases of possible combinations of the resonant radii of the string loop oscillations. At each of the three observed events, and each of the resonant points, the fitting is related to the upper of the observed frequencies (or the common frequency at $r_{1:1}$); the precision of their measurement is also taken into account. The fitting procedure then gives for each of the observed events an allowed region of the parameter space of the spacetime parameters $M,a$, determined by the limiting values of the string loop parameter $\omega \in \langle-1,1\rangle$. Due to the degeneracy of the radial profiles of the string loop oscillation frequencies in the \Schw{} spacetimes ($a=0$), i.e., their independence of the stringy parameter $\omega$, the fitting predicts only one value of the mass parameter $M$ for the spin $a=0$ at each observational event. Extension of the allowed region related to the whole interval of string loop parameter $\omega \in \langle-1,1\rangle $ (i.e., the interval of allowed values of $M$) increases with increasing spin $a$. Therefore, the string loop oscillation model implies a "triangular" limit on the spacetime parameters $M,a$ for each of the observed events of HF QPOs -- see Figure \ref{fits1}. In a given spacetime, the limits have to be satisfied simultaneously and we directly obtain the region of allowed values of the spacetime parameters, if the theoretical restrictions on the mass ($M < 2.5 {\rm M}_{\odot}$) and spin ($a < 0.4$ related to the Hartle-Thorne model of neutron stars) are also taken into account. (For completeness, we demonstrate in Figure \ref{fits1} also the limits implied by the general restriction on the neutron star spin, $a < 0.7$). Along with the restrictions on the spacetime parameters $M$ and $a$, we  obtain simultaneously restrictions on the stringy parameter $\omega$. Moreover, the restrictions on the radii where the resonant oscillations occur have to be also taken into the account, if we test the Hartle-Thorne models of the neutron star for concrete equations of state. The results representing the limits on the spacetime and string loop parameters $M,\omega$ are presented in Table 2 for the characteristic limiting values of the spin parameter $a$. 

\begin{figure}
\centering 
\includegraphics[height=\picB]{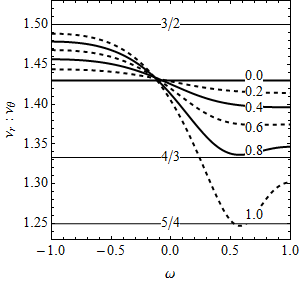}
\caption{ \label{fexample}
Ratio of frequencies $\nu_\mir : \nu_\mit$ calculated at the local maximum of $\nu_\mir(r)$ function determining the frequency of the radial oscillations of the string loops for different values of parameter $a$.}
\end{figure}

Our results indicate that only the case 4, with both the resonant radii related to $r_{2:3}$, is physically realistic, as the other three cases, where the radius $r_{3:2}$ enters the play, give unrealistic values of the mass and spin of the neutron star, and, moreover, the radius $r_{3:2}$ could be too close to the surface of the neutron star. Using the results obtained in the case 4, we can see that for the lowest value of the spin $a=0.2$, the unique value of the neutron star mass reads $M=2.35 {\rm M}_{\odot}$. For the allowed spin $a=0.3$, the mass parameter can be in the interval $M \sim (2.3 - 2.4) {\rm M}_{\odot}$, while for the upper limit on the spin of the source $a=0.4$, the mass parameter can be in the interval $M \sim (2.2 - 2.5){\rm M}_{\odot}$. Such large predicted values of the neutron star mass are in agreement with the assumption of applicability of the Kerr geometry in our study. Further, we obtained also restrictions on the stringy parameter $\omega$ that has to be positive at the resonant radii $r_{1:1}$, but negative at the resonant radii $r_{2:3}$. 

In Figure \ref{rprofile} we represent the situation of fitting the data from the three observational events in the case 4 for some realistic values of the spacetime and stringy parameters from the allowed interval of the parameters. Notice that the resonant points with the frequency ratio $2:3$ are located in vicinity of the local maximum of the radial profile of the radial frequency. It means that the upper frequency can be nearly the same for a wide range of radii of the stable equilibria fulfilling the resonance condition at $r_{2:3}$. The resonance can be thus obtained by slight shift in the radius, since the upper (radial) frequency remains nearly constant at vicinity of the local maximum of the radial frequency radial profile, while the vertical frequency varies strongly in this region. The situation is illustrated by Figure \ref{fexample} demonstrating that the frequency ratio $\nu_{\mir}:\nu_{\mit}$ is relatively close to $3:2$ at the local maximum of $\nu_{\mir}(r)$ for negative values of the stringy parameter $\omega$, but the ratio decreases for positive values of $\omega$. The string loop oscillation model predicts negative values of the parameter $\omega$ at the resonance radii $r_{2:3}$ implying thus their occurrence in the vicinity of the local maximum of $\nu_{\mir}(r)$.

\section{Conclusions}

We have tested the string loop oscillation model introduced in \cite{Stu-Kol:2014:PHYSR4:} by fitting the frequencies of the three HF QPOs observed in the XTE~J1701-407 LMXB source containing a neutron star, using the analytic formulae for the radial and vertical oscillations in the Kerr spacetime approximating the external geometry of the neutron star. This model, reflecting oscillations of the string loops governed by interplay of tension and angular momentum, gives relevant restrictions on the spacetime parameters $M,a$ and the string loop parameter $\omega$ that must be varied for the three observational events. The restricting results can be summarized as follows:
\begin{itemize}
\item we cannot fit the observed data assuming only one string loop having a fixed value of the parameter $\omega$,
\item the neutron star has to be rather fast rotating, having the dimensionless spin limited to the interval $0.2 < a < 0.4$,
\item the neutron star has to be very massive, with mass parameter limited to the interval  $2.1\mathrm{M_{\odot}} < M < 2.5\mathrm{M_{\odot}}$.
\end{itemize}

Since the XTE~J1701-407 neutron star has to be very massive, we can conclude that the application of the Kerr geometry in the fitting procedure is justified, as for the near-maximum-mass neutron stars the exterior Hartle-Thorne geometry has to be close to the exterior Kerr geometry, giving close predictions of the physical phenomena occurring in their vicinity. Of course, we have to test in future, if application of the string loop model related to the Hartle-Thorne geometry with large values of $q/a^2 > 3$, when complex behaviour of the frequency relations is expected \cite{Tor-etal:2014:ASTRA:}, could allow for fitting the data by neutron star models with intermediate or low mass, $M < 2{\rm M}_{\odot}$. 

In a future study, we would like to test the string loop oscillation model, potentially reflecting  tension of magnetic field lines in the oscillating objects, also for the HF QPOs observed in typical atoll sources demonstrating large scatter of the observed twin HF QPOs. 

We can conclude that the string loop oscillation model of the HF QPOs occurring in the strong gravitational field of compact objects can fit both the high, kHz frequencies observed in the microquasars \cite{Stu-Kol:2014:PHYSR4:}, i.e., binary systems containing a black hole, and the kHz frequencies observed in the special, low-luminosity atoll binary system XTE J1701-407 containing a neutron star and demonstrating a special set of HF QPOs untypical for the atoll sources. Therefore, it is worth to investigate the string loop oscillation model in more detailed way. First, we have to concentrate on the optical phenomena related to the radiation of the oscillating string loops that could be treated in the standard way introduced for the electromagnetic (gravitational) radiation of strings \cite{Vil-She:1994:CSTD:} and modification of the emitted radiation by strong gravity of a black hole or a neutron star. Second, we have to study the conditions for creation of the "magnetic" string loops along the lines proposed in \cite{Cre-Stu:2013:PHYSRE:,Cre-Stu-Tes:2013:PlasmaPhys:}, giving the relations of the physical parameters of magnetized plasma to the stringy angular momentum parameters $J$ and $\omega$.  

\begin{acknowledgements}
The authors acknowledge the Albert Einstein Centre for Gravitation and Astrophysics supported by the~Czech Science Foundation Grant No. 14-37086G. MK would like to thank the internal student grant SGS/23/2013 of the Silesian University. 
\end{acknowledgements}



\def\prc{Phys. Rev. C}
\def\pre{Phys. Rev. E}
\def\prd{Phys. Rev. D}
\def\jcap{Journal of Cosmology and Astroparticle Physics}
\def\apss{Astrophysics and Space Science}
\def\mnras{Monthly Notices of the Royal Astronomical Society}
\def\apj{The Astrophysical Journal}
\def\aap{Astronomy and Astrophysics}
\def\actaa{Acta Astronomica}
\def\pasj{Publications of the Astronomical Society of Japan}
\def\apjl{Astrophysical Journal Letters}
\def\pasa{Publications Astronomical Society of Australia}
\def\nat{Nature}
\def\physrep{Physics Reports}
\def\araa{Annual Review of Astronomy and Astrophysics}
\def\apjs{The Astrophysical Journal Supplement}


\end{document}